# Superconductivity at 44.4 K achieved by intercalating EMIM$^+$ into FeSe[*]

Jinhua Wang (王晋花)[1], Qing Li (李庆)[1], Wei Xie (谢威)[1], Guanyu Chen (陈冠宇)[1], Xiyu Zhu(祝熙宇)[1†] and Hai-Hu Wen (闻海虎)[1‡]

[1]*Center for Superconducting Physics and Materials, National Laboratory of Solid State Microstructures and Department of Physics, Collaborative Innovation Center of Advanced Microstructures, Nanjing University, Nanjing 210093, China*

Superconductivity with transition temperature $T_c$ above 40 K was observed in protonated FeSe (H$_y$-FeSe) previously with the ionic liquid of EMIM-BF$_4$ used in the electrochemical process. However, the real superconducting phase is not clear until now, and detailed structural, magnetization and electrical transport measurements are lacking. By using similar protonating technique on FeSe single crystals, we obtained superconducting samples with $T_c$ above 40 K. We show that the obtained superconducting phase is not H$_y$-FeSe but actually an organic-ion (C$_6$H$_{11}$N$_2^+$ referred to as EMIM$^+$)-intercalated phase (EMIM)$_x$FeSe. By using X-ray diffraction technique, two sets of index peaks corresponding with different $c$-axis lattice constants are detected in the protonated samples, which belong to the newly formed phase of intercalated (EMIM)$_x$FeSe and the residual FeSe, respectively. The superconductivity of (EMIM)$_x$FeSe with $T_c$ of 44.4 K is confirmed by resistivity and magnetic susceptibility measurements. Temperature dependence of resistivity with different applied magnetic fields reveals that the upper critical field $H_{c2}$ is quite high, while the irreversibility field $H_{irr}$ is suppressed quickly with increasing temperature till about 20 K. This indicates that the resultant compound has a high anisotropy with a large spacing between the FeSe layers.

**Keywords:** FeSe, iron-based superconductor, electrochemical intercalation

**PACS:** 74.70.Xa, 74.25.-q, 74.72.Ek, 82.45.Aa

## 1. Introduction

Iron-based superconductors have attracted vast interest in condensed matter physics and material science since it was discovered in 2008.[1] Among iron-based superconducting materials, the compound FeSe was reported to have the simplest structure with a critical transition temperature ($T_c$) of about 8 K at ambient pressure,[2] which is constituted by edge-sharing FeSe$_4$-tetrahedra layers stacking along the $c$-axis. By applying high pressures, a dome-shaped superconducting region with rich physics appears, and its $T_c$ was enhanced up to 36.7 K at 8.9 GPa.[3-4] Due to the lack of charge carrier reservoir layers, it is natural to dope carriers into the FeSe planes by intercalation in order to increase its transition temperature. Intercalating guest elements or composites into adjacent FeSe layers is the most common way to achieve that goal.

[*] Project supported by the National Natural Science Foundation of China (Grant No. 12061131001 and No. 52072170), and the Strategic Priority Research Program of Chinese Academy of Sciences (Grant No. XDB25000000).
[†] Corresponding author. E-mail: zhuxiyu@nju.edu.cn
[‡] Corresponding author. E-mail: hhwen@nju.edu.cn

In the beginning, the alkali metals were intercalated into FeSe by solid state reactions, through which $A_x$Fe$_{2-y}$Se$_2$ ($A$ = K, Rb, Cs, Tl, x<1, y<1) was synthesized with $T_c$ around 32 K.[5-9] However, in order to satisfy the charge balance, phase separation occurs in the body of the material, leading to the coexistence of superconducting phase $A_x$Fe$_{2-y}$Se$_2$ and antiferromagnetic insulating phase $A_2$Fe$_4$Se$_5$.[10-13] This prohibits a thorough and systematic investigation of the physical properties in those materials. Thus, it is necessary to have a low-temperature technique to prepare intercalated FeSe-layer materials. Due to the special character of dissolving multiple metals, liquid ammonia could help to insert not only alkali metals (Li, Na, K, etc.), but also alkali-earth metals (Ca, Sr, Ba) and rare-earth metals (Eu, Yb) into the FeSe bulk samples.[14-17] These metals are co-inserted with liquid ammonia molecules, thus the inserted molecules are closer to be neutral in charge comparing with the alkali metal ion with valence state of +1, the former causes relatively complete FeSe planes with a significant increase of the $c$-axis lattice constant. And the highest $T_c$ of 46 K was reached among this series of superconductors.[14] Although a large increase of $T_c$ was realized, the chemical activity of intercalating composites between the adjacent FeSe layers made those materials extremely unstable in air. Using the hydrothermal technique and improved hydrothermal ion-exchange process,[18-20] a stable compound Li$_{1-x}$Fe$_x$OHFeSe with an ordered guest-layer Li$_{1-x}$Fe$_x$OH was prepared with $T_c$ up to 42 K.[20] Besides the inorganic molecules which are co-inserted with metals, organic molecules could also be intercalated into FeSe,[21-28] producing a series of superconductors with different $c$-axis lattice parameters. Due to the varying size of organic molecules, such as C$_5$H$_5$N, (H$_2$N)C$_n$H$_{2n}$(NH$_2$), and C$_n$H$_{2n+3}$N ($n$ = 6, 8, 18),[21, 24, 28] the interlayer spacing could be largely stretched to a certain degree, and the largest $c$-axis lattice parameter of 55.7 Å was achieved in Li$_x$(ODA)$_y$Fe$_{1-z}$Se with $T_c$ of about 42 K.[28]

Besides plenty of researches about alkali metals co-inserted with inorganic or organic molecules into FeSe, some new superconductors were also discovered simply by intercalating organic composites, forming for example (C$_2$H$_8$N$_2$)$_x$FeSe.[29] This may pave a new way to synthesize intercalated FeSe derivative superconductors. The intercalation of C$_2$H$_8$N$_2$ in FeSe made $c$-axis lattice parameter expanded up to 21.700(6) Å. Moreover, two other different organic ions, cetyltrimethyl ammonium (CTA$^+$) and tetrabutyl ammonium (TBA$^+$),[30-31] were successfully inserted into FeSe through electrochemical intercalation. It has been found that these two kinds of organic-ions intergrown with FeSe formed a bulk superconductivity showing $T_c$ of 45 K and 50 K, with $c$-axis lattice parameter expanded up to 14.5 Å and 15.5 Å, respectively. Through the similar method, a new derivative of FeSe, the so called protonated FeSe (H$_y$-FeSe), was discovered with $T_c$ of 41 K,[32] in which the existence of hydrogen was indicated by nuclear magnetic resonance (NMR) measurements. But it is still unclear what is the real superconducting phase, and studies on the detailed structure and physical properties are still lacking up to now.

In this work, we give detailed investigation on the above material, named as protonated FeSe, obtained through the same method and interactant, 1-ethyl-3-methylimidazolium tetrafluoroborate (EMIM-BF$_4$), as reported by Cui *et al*.[32] According to the X-ray diffraction (XRD) measurements, we show that the title compound has a highly expanded *c*-axis constant, thus we conclude that the relevant phase is not H$_y$-FeSe, but rather another organic-ion-intercalated FeSe [(EMIM)$_x$FeSe]. The $T_c$ of this new FeSe-based superconductor is 44.4 K, as evidenced by the results of temperature dependent magnetic susceptibility and resistivity measurements. Note that our intercalated (EMIM)$_x$FeSe samples are also not very stable in atmosphere, very similar to the (TBA)$_{0.3}$FeSe,[31] and after a few days it will degrade back to the pristine FeSe with $T_c$ of about 8 K.

## 2. Experimental details

The schematic experimental setup of the electrochemical intercalation experiment is illustrated in Fig. 1(a).[32] As shown in the illustration, the positive and negative electrodes made of platinum are placed in an ionic-liquid container. Then this device is put into a heating mantle to maintain a certain temperature during the electrochemical process. The FeSe single crystal grown by means of chemical vapor transport technique is attached on the negatively charged electrode by silver paste,[33-34] while positive electrode Pt wire is placed right opposite to it. After loading the FeSe single crystal on the electrode, we add ionic liquid into the container till the sample is completely immersed. During the electrochemical process, a constant voltage of 4 V is applied and the temperature of the heating mantle is set to 355 K. After about three days of electrochemical reaction, the intercalated FeSe crystal is removed from the electrode and various measurements are taken after cleaning the silver paste and some residual reactants on the sample surface.

The XRD is conducted on a Bruker D8 Advanced diffractometer with the Cu $K_\alpha$ radiation at room temperature. The temperature dependent resistivity is measured by a physical property measurement system (PPMS-16 T, Quantum Design) through the typical four-probe method at different magnetic fields. The temperature dependent magnetic susceptibility is carried out by Quantum Design PPMS with Vibrating Sample Magnetometer (VSM), while magnetic field is applied parallel to the *c*-axis of the sample.

## 3. Results and discussion

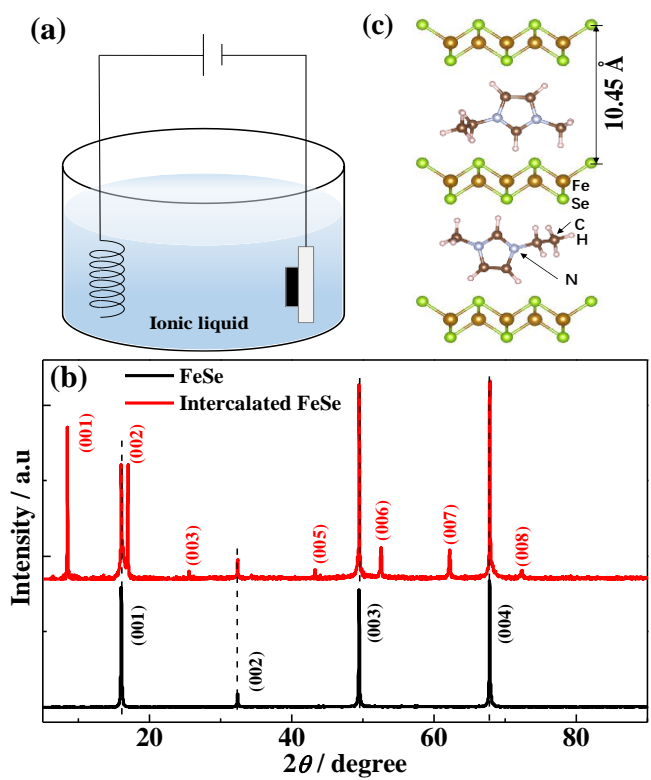

**Fig. 1.** (Color line) (a) An illustration of the device for the electrochemical reaction. (b) X-ray diffraction patterns of FeSe single crystal and intercalated FeSe. The Miller indices of the intercalated phase is colored by red, and that of FeSe phase is colored by black, respectively. (c) The schematic crystal structure of (EMIM)$_x$FeSe.

Figure 1(b) shows the XRD patterns of pristine FeSe single crystal and intercalated FeSe after the electrochemical reaction. No matter before or after the experiment, the sample remains in having good crystallinity. Hence, the XRD patterns show an obvious *c*-axis orientation and all peaks are indexed perfectly as (*00l*) on the basis of tetrahedral structure. The black solid line represents a series of (*00l*) peaks of FeSe single crystal, which indicates a *c*-axis lattice parameter of 5.52 Å. Meanwhile, the red solid line contains two sets of peaks. One set of them shows the same diffraction peaks as the black line, belonging to the residual FeSe of sample. The other set represents a new phase, in which the *c*-axis lattice constant is found to be 10.45 Å taking the space group of P4/nmm. Such an obvious increase of *c*-axis lattice parameter suggests that some kind of molecules with big size in the experiment has been inserted into the sample and the spacing between adjacent FeSe layers is expanded. Considering the fact that the pristine FeSe is held on the negative electrode, we intend to conclude that the inserted molecules are most likely organic ions EMIM$^+$ from the ionic liquid. The schematic structure of this new phase is shown in Fig. 1(c). Considering the finite size for three dimensions of organic ions, they probably arrange themselves with a specific orientation between the adjacent FeSe layers as shown in Fig. 1(c). However, this needs further verification by determining the internal structure.

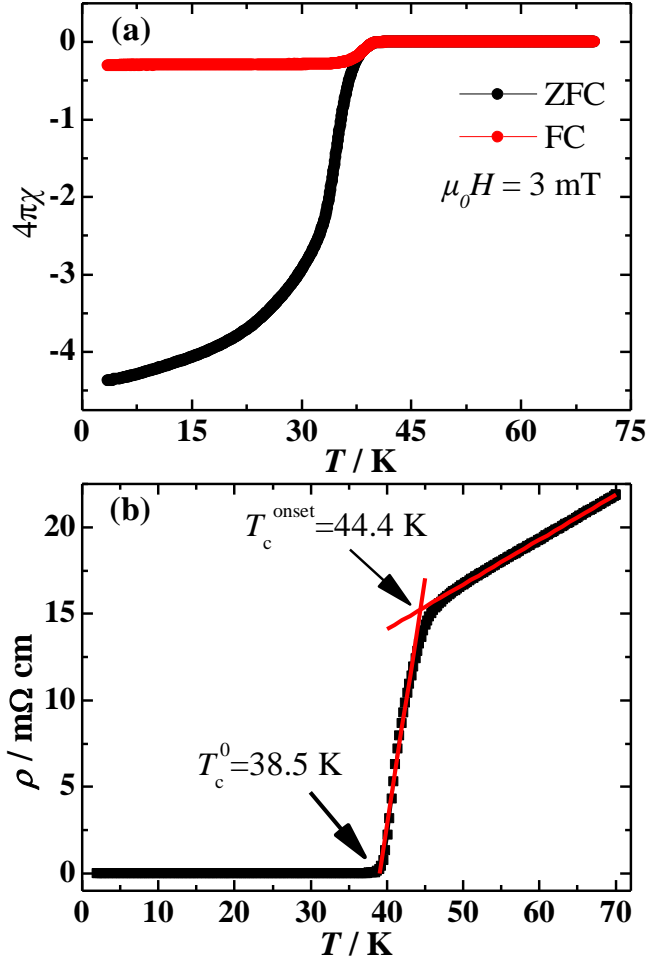

**Fig. 2.** (Color line) (a) Temperature dependent magnetic susceptibility of intercalated FeSe single crystal measured in ZFC and FC modes under a magnetic field of 3 mT. (b) Temperature dependent resistivity under zero magnetic field.

The temperature dependence of magnetic susceptibility and resistivity have been measured, and the results are shown in Fig. 2. It is worth mentioning that two samples from the same batch were used for the magnetic and transport measurements since the quality of the samples degenerates quickly in the transfer process between two types of measurements. Figure 2(a) exhibits the temperature dependent magnetic susceptibility ($\chi$-T curve) of intercalated FeSe, which is measured in zero-field-cooled (ZFC) and field-cooled (FC) modes with an applied magnetic field of 3 mT. The magnetic screening volume calculated from the ZFC data is about 437% comparable to the value reported previously in the similar systems,[30-32] which is larger than 100% as the demagnetization effect has not been taken into account. A sharp transition shows up at about 40 K in the $\chi$-T curve, demonstrating the emergence of superconductivity. Note that the $T_c$ of FeSe is around 8 K,[2] but this superconducting transition is not visible in the $\chi$-T curve. The contradiction between XRD pattern and $\chi$-T curve may be explained in the following way. The intercalation proceeds from the outside to the inside of sample, thus the surface of sample is intercalated successfully and uniformly while there is still residual FeSe in the interior of the sample. Figure 2(b) presents the temperature dependent resistivity under zero magnetic field with a current of 100 μA. There is a dramatic decrease of resistivity at 44.4 K, which is roughly consistent with the $T_c$ obtained from the $\chi$-T curve. The resistivity reaches zero at 38.5 K.

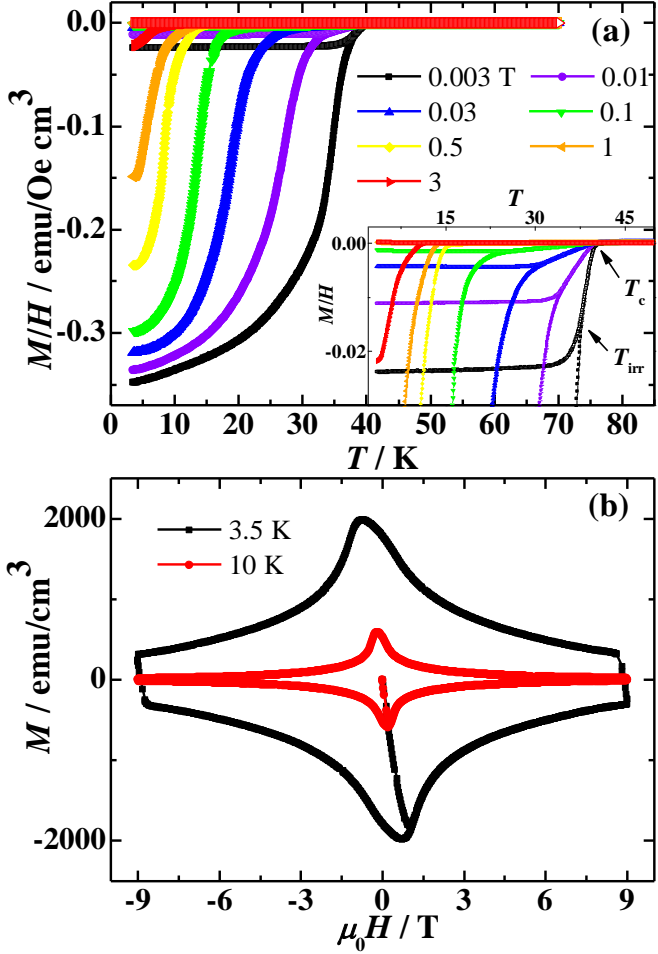

**Fig. 3.** (Color line) (a) Temperature dependent magnetic susceptibility of intercalated FeSe single crystal measured at different magnetic fields. Inset shows the enlarged view of the *M-T* curves near the transition. (b) Magnetization hysteresis loops (MHLs) of intercalated FeSe single crystal at 3.5 and 10 K.

In order to check the magnetic characteristics of the intercalated crystals, we have measured the temperature dependent magnetic susceptibility of intercalated FeSe in ZFC and FC modes at different fields (*M-T* curves), and the results are shown in Fig. 3(a). At low fields, the ZFC curves show a trend of saturation. With increasing of external magnetic fields, the magnetization value of ZFC curve at low temperature was suppressed, which is corresponding to the decrease of magnetic screening volume. The inset of Fig. 3(a) shows the enlarged view of *M-T* curves near the transition. The deviation point of the ZFC and FC curves can be defined as the irreversible temperature $T_{irr}$ at the external fields, and the $T_{irr}$ decreases quickly with increasing applied magnetic field. The temperature dependence of irreversibility field $H_{irr}(T)$ (*M-T*) of intercalated FeSe is shown in Fig. 4(b). Figure 3(b) shows the magnetization hysteresis loops (MHLs) at 3.5 and 10 K, respectively. The MHLs show a typical magnetic hysteresis behavior of type-II superconductors. The width of MHL measured at 3.5 K is much wider than that at 10 K, and the $\Delta M$ of MHLs at low temperatures are comparable to other iron-based superconductors, like $(Li_{1-x}Fe_x)OHFeSe$,[35] which indicates the good vortex pinning of our samples.

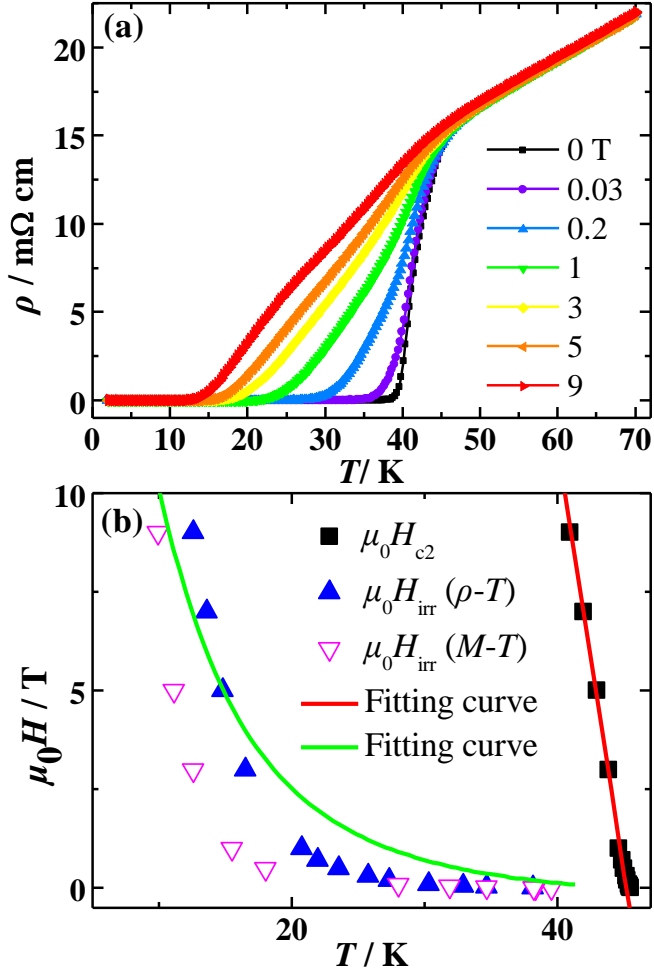

**Fig. 4.** (Color line) (a) Temperature dependent resistivity under different applied magnetic fields. (b) $H$-$T$ phase diagram of the intercalated FeSe. Black squares represent the upper critical field $H_{c2}(T)$ data. Blue (magenta) symbols represent the irreversibility field $H_{irr}(T)$ obtained from the temperature dependent resistivity (magnetic susceptibility) data. The red and green solid lines show the fitting results of $H_{c2}(T)$ and $H_{irr}(T)$, respectively.

Figure 4(a) displays the temperature dependent resistivity of another intercalated FeSe sample under different applied magnetic fields with a standard four-probe method. As we can see, the onset transition temperature $T_c$ of intercalated FeSe is rather stable under external magnetic fields. In contrast to this, the zero-resistance temperature $T_c$ decrease very fast with increasing magnetic fields, which is analogous to the situation of $M$-$T$ curves. Figure 4(b) gives the $H$-$T$ phase diagram of the intercalated FeSe. The upper critical field $H_{c2}$ and the irreversibility field $H_{irr}$ as a function of temperature are obtained from the resistivity and magnetic susceptibility data. The $H_{c2}$ is obtained by using the criterion of $95\%\rho_n(T)$ in $\rho$-$T$ curves, and the $H_{irr}(\rho$-$T)$ is determined by using the criterion of $0.2\%\rho_n(T)$, respectively. It can be seen that, the $H_{c2}$ shows a rather steep increase with lowering temperature, indicating a quite high upper critical field at zero temperature. The temperature dependence of $\mu_0 H_{c2}(T)$ exhibits an almost linear behavior near $T_c$, which gives rise to a slope of $d\mu_0 H_{c2}/dT = -2.17$ T/K. Thus the upper critical field at zero temperature calculated by the Werthamer-Helfand-Hohenberg (WHH) formula $\mu_0 H_{c2}(0) = 0.69 T_c |d\mu_0 H_{c2}/dT|_{T_c}$ is about 66.48 T.[36]

According to the Ginzburg-Landau theory and Pippard formula, it is known that $\mu_0 H_{c2} \approx (\pi \Phi_0 / 2\hbar^2)(\Delta/v_F)^2$ with $\Phi_0$ the flux quanta, $\Delta$ the superconducting gap, $v_F$ the Fermi velocity. Thus, high value of $H_{c2}(0)$ reflects a strong pairing gap. On the contrary, the irreversibility field $H_{irr}$ decreases quickly with increasing temperature till about 20 K. Above 20 K, the $H_{irr}$ drops slowly, leading to a wide region between the $H_{irr}(T)$ and $H_{c2}(T)$. It is known that above $H_{irr}(T)$ there is finite dissipation, thus a wider region of vortex liquid is observed comparing with the original FeSe.[37] This broadening of superconducting transition is similar to other two-dimensional superconductors, such as Li$_{1-x}$Fe$_x$OHFeSe,[35] which suggests a large spatial distance between FeSe layers and a quite high anisotropy of intercalated FeSe samples. And then the $H_{irr}(\rho\text{-}T)$ is fitted by the melting line of flux line lattice (FLL):[38]

$$\frac{t}{(1-t)^{1/2}} \frac{b^{1/2}}{1-b} \left( \frac{4(\sqrt{2}-1)}{(1-b)^{1/2}} + 1 \right) = \alpha \quad (1)$$

with $t = T/T_c$, $b = H/H_{c2}$, which is deduced from the Lindermann criterion.[39] The parameter $\alpha$ is given by the formula: $\alpha = 2\pi(\epsilon M_z/M)^{-1/2} c^2$, with the Lindermann number $c \approx 0.15 - 0.25$ and $\epsilon = 16\pi^3 \kappa^4 (k_B T_c)^2 / \Phi_0^3 H_{c2}^0$, in which $\kappa$ is the Ginzburg-Landau parameter. The mass ratio $(M_z/M)^{1/2}$ becomes lager as the spacing distance between FeSe layers increases, where $M_z$ is a quasiparticle effective mass along the $c$ axis and $M$ describes the mass in the FeSe planes. The optimal result of $\alpha = 0.33$ is obtained by fitting $H_{irr}(T)$ data. In the fitting process, the parameter of $\kappa$ in high-$T_c$ superconductors usually taken as a large value like $\kappa = 100$, such as when it equals to 95, the $H_{irr}(T)$ data of Bi$_{2.2}$Sr$_2$Ca$_{0.8}$Cu$_2$O$_8$ could be well fitted by the melting criterion.[38] Thus, we take a general value for $\kappa = 100$ as the intercalated FeSe is high-$T_c$ superconductor ($T_c$ = 44.4 K). By substituting the $H_{c2}(0)$, $T_c$ and $c = 0.2$ into the formula of $\alpha$, we get the mass ratio $(M_z/M)^{1/2} \approx 44.5$, which is comparable with the value of a quasi-two-dimensional (Quasi-2D) superconductor Bi$_{2.2}$Sr$_2$Ca$_{0.8}$Cu$_2$O$_8$ $[(M_z/M)^{1/2} = 60]$,[38] further proving the high spatial anisotropy of the sample.

## 4. Conclusions

We report superconductivity with transition temperature $T_c$ above 40 K by electrochemically reacting FeSe with the ionic liquid of EMIM-BF$_4$. This phenomenon was recently reported as a process of protonating the FeSe crystals, but it was not clear what is the real superconducting phase. By doing X-ray diffraction measurements, we found a new set of diffraction peaks arising from the intercalated samples, these newly emergent peaks coexist with those of residual FeSe. We conclude that the newly emergent superconducting phase is the organic-ion (C$_6$H$_{11}$N$_2^+$, EMIM$^+$)-intercalated FeSe phase, namely (EMIM)$_x$FeSe. Our results rule out the possibility that the high temperature superconducting phase is the simple protonated one H$_y$-FeSe. We also find that the upper critical field $H_{c2}$ is quite high, indicating a strong pairing potential. While the irreversibility field $H_{irr}$ is suppressed quickly with increasing temperature showing a large region of vortex liquid in the phase diagram. By fitting the irreversibility line to the vortex melting formula deduced from the Lindermann criterion, we get a quite large mass ratio $(M_z/M)^{1/2} \approx 44.5$, which is comparable to the value of a Quasi-2D high-$T_c$ superconductor $Bi_{2.2}Sr_2Ca_{0.8}Cu_2O_8$. This also supports that the resultant material should be highly anisotropic, as expected for a system with a large spacing between the FeSe layers.

**Acknowledgment**